\documentclass[aps,pra,twocolumn,letterpaper,superscriptaddress,10pt]{revtex4-1}
\usepackage{amsmath,amssymb}
\usepackage{graphicx,mathptmx,ulem}
\usepackage[dvipsnames]{xcolor}
\usepackage[colorlinks=true,urlcolor=blue,citecolor=blue,linkcolor=magenta]{hyperref}

\begin{document}

\title{Quantum Correlations beyond Entanglement and Discord}

\author{S. K\"ohnke}
	\email{semjon.koehnke@uni-rostock.de}
	\affiliation{Arbeitsgruppe Experimentelle Quantenoptik, Institut f\"ur Physik, Universit\"at Rostock, D-18051 Rostock, Germany}

\author{E. Agudelo}
	\affiliation{Institute for Quantum Optics and Quantum Information --- IQOQI Vienna, Austrian Academy of Sciences, Boltzmanngasse 3, 1090 Vienna, Austria}

\author{M. Sch\"unemann}
	\affiliation{Arbeitsgruppe Experimentelle Quantenoptik, Institut f\"ur Physik, Universit\"at Rostock, D-18051 Rostock, Germany}
	\affiliation{Department of Ophthalmology, Rostock University Medical Center, Rostock, Germany}
    \affiliation{Department Life, Light and Matter, University of Rostock, Rostock, Germany}

\author{O. Schlettwein}
	\affiliation{Arbeitsgruppe Experimentelle Quantenoptik, Institut f\"ur Physik, Universit\"at Rostock, D-18051 Rostock, Germany}

\author{W. Vogel}
	\affiliation{Institut f\"ur Physik, Universit\"at Rostock, D-18051 Rostock, Germany}

\author{J. Sperling}
    \email{jan.sperling@upb.de}
	\affiliation{Integrated Quantum Optics Group, Institute for Photonic Quantum Systems (PhoQS), Paderborn University, Warburger Stra\ss{}e 100, 33098 Paderborn, Germany}

\author{B. Hage}
	\affiliation{Arbeitsgruppe Experimentelle Quantenoptik, Institut f\"ur Physik, Universit\"at Rostock, D-18051 Rostock, Germany}
    \affiliation{Department Life, Light \& Matter, University of Rostock, Rostock, Germany}

\begin{abstract}
	Dissimilar notions of quantum correlations have been established, each being motivated through particular applications in quantum information science and each competing for being recognized as the most relevant measure of quantumness.
	In this contribution, we experimentally realize a form of quantum correlation that exists even in the absence of entanglement and discord.
	We certify the presence of such quantum correlations via negativities in the regularized two-mode Glauber-Sudarshan function.
	Our data show compatibility with an incoherent mixture of orthonormal photon-number states, ruling out quantum coherence and other kinds of quantum resources.
	By construction, the quantumness of our state is robust against dephasing, thus requiring fewer experimental resources to ensure stability.
	In addition, we theoretically show how multimode entanglement can be activated based on the generated, nonentangled state.
	Therefore, we implement a robust kind of nonclassical photon-photon correlated state with useful applications in quantum information processing.
\end{abstract}

\date{\today}
\maketitle

\section{Introduction}
	The certification of quantum correlations is essential for the ever-accelerating development of quantum technologies.
	Beyond this practical demand, a fundamental understanding of quantum correlations, including their characterization and quantification, plays a key role when exploring the boundary between classical theories and the unique features of quantum physics.
	Still, the question remains which kinds of correlation are genuinely quantum.
	That is, which of the many contenders---be it an established or recently proposed concept (e.g., quantum coherence and resource theory \cite{SAP17,CG19}, entanglement \cite{HHHH09}, discord \cite{MBCPV12}, etc.)---does describe the concept of a nonclassical correlation best?
	In this work, we show that the notion of nonclassicality in quantum optics \cite{TG65,M86} can supersede contemporary forms of quantumness in its ability to unveil quantum correlations.

	In the context of quantum optics, any observations which cannot be fully described in terms of Maxwell's wave theory of light are called nonclassical \cite{TG65,M86}.
	This well-established concept of nonclassicality is based on the impossibility of describing field correlations as done in classical electrodynamics, and it is commonly defined in terms of the Glauber-Sudarshan $P$ representation \cite{G63,S63}.
	The latter describes nonclassical light through phase-space distributions that are, in the case of quantum light, incompatible with a classical concept of a non-negative probability distribution.

	The more recently developed concept of quantum coherence adapts some ideas of the notion of nonclassicality to quantify resources required for quantum information processing  \cite{SAP17,CG19}.
	In this framework, quantum superpositions equally serve as the origin of quantumness in a system.
	However, in most cases, the classical reference is defined through incoherent mixtures of orthonormal basis states, contrasting the notion of quantum-optical nonclassicality in terms of nonorthogonal eigenstates of the non-Hermitian annihilation operator.

	Entanglement, which can be embedded into the concept of coherence \cite{SSDBA15,KSP16}, is by far the most frequently studied form of quantum correlation among the many contenders \cite{HHHH09}.
	This is due to its fundamental role as well as its many applications, e.g., in quantum metrology, cryptography, computing, and teleportation.
	The phenomenon of entanglement was discovered in early seminal discussions about the implications of quantum physics \cite{S35,EPR35}, long before the conception of the relatively young field of quantum information processing.

	Many other notions and measures of quantum correlations have been proposed too \cite{MBCPV12}.
	For instance, discord is a feature which includes correlations caused by entangled but also by nonentangled states \cite{HV01,OZ01}, and it can be connected to quantum coherence as well \cite{MYGVG16}.
	In this context, it is worth mentioning that the label quantum for this sort of correlation is a topic of ongoing debates \cite{GOS15}.
	Nonetheless, it has been demonstrated that discord is maximally inequivalent to the notion of quantum-optical nonclassicality \cite{FP12}.
	To date, it remains an open problem to decide---not only in theory, but also experimentally---which of the candidates is best suited for characterizing quantum correlations.

	In this Letter, we address this issue experimentally by realizing and analyzing a fully phase-randomized two-mode squeezed vacuum (TMSV) state \cite{ASV13}.
	This state of quantum light has the following properties:
        It is nonentangled;
        it has zero discord;
        it does not exhibit quantum coherence in the photon-number basis;
        its reduced single-mode states are classical;
        and it has a non-negative, two-mode Wigner function.
	Despite these strong signatures of classicality, we demonstrate the presence of quantum correlations as defined through the notion of nonclassicality in quantum optics with a statistical significance next to certainty.
	Because of the phase independence of the generated state, this nonclassical feature is robust against dephasing.
	Furthermore, the activation of entanglement using this kind of state is developed to show its resourcefulness for quantum information processing applications.

\section{Quantum correlations}
	For analyzing quantum correlations, we consider a class of two-mode states that are phase insensitive \cite{FP12,ASV13}.
	Still, intensity-intensity (likewise, photon-photon) correlations are present in such states.
	For instance, this can be achieved by a full phase randomization of a TMSV state, resulting in
	\begin{align}
        \label{eq:theo}
		\hat\rho=\sum_{n\in\mathbb N} (1-p)p^n |n\rangle\langle n|\otimes|n\rangle\langle n|,
	\end{align}
	where $p=\tanh|\xi|$ is a value between zero and one and $\xi$ is the complex squeezing parameter.
	Such states are, for example, a relevant resource for boson sampling tasks \cite{SLR17}.

	In terms of quantum correlations, it can be directly observed that the state in Eq. \eqref{eq:theo} is an incoherent mixture of photon-number states, thus exhibiting no quantum coherence in the form of quantum superpositions of photon-number states;
	it is a classical mixture of tensor-product states, thus exhibiting no entanglement;
	and it has zero discord because $\hat\rho=\sum_{n\in\mathbb N}\hat\rho_{A|n}\otimes |n\rangle\langle n|$ holds true, where the photon-number states form the eigenbasis to $\hat\rho_{A|n}=(\hat 1\otimes \langle n|)\hat\rho(\hat 1\otimes|n\rangle)$ and $\mathrm{tr}_A\hat\rho$ \cite{D11}.
	For those criteria of quantum correlations, it suffices in our scenario to consider the contribution of off-diagonal elements,
	\begin{align}
		\label{eq:CohMeasure}
		\mathcal C(\hat\rho)\stackrel{\text{def.}}{=}\sum_{\substack{m,n,k,l\in\mathbb N: \\ m\neq n, k\neq l}}
		\big|(\langle m| \otimes \langle k|)\hat\rho(|n\rangle\otimes |l\rangle)\big|,
	\end{align}
	which quantifies the coherent contributions \cite{BCP14} and is $\mathcal C(\hat\rho)=0$ for the state in Eq. \eqref{eq:theo}.
	It is worth emphasizing that the nullity of coherence in the two-mode photon-number basis implies the nullity of discord which further implies no entanglement.
	In addition, the incoherent mixture of photon-number states under study further implies a classical interpretation in a particle picture \cite{SDNTBBS19}.
	Furthermore, the marginal states $\mathrm{tr}_A\hat\rho$ and $\mathrm{tr}_B\hat\rho$ are thermal states and, thus, classical, too.
	Also, the two-mode state under study is a mixture of Gaussian TMSV states, implying a non-negative Wigner function.

	At this point, one sees no indication of quantum correlations.
	Yet, we have not considered the notion of nonclassicality in quantum optics so far.
	This concept is defined through the Glauber-Sudarshan $P$ representation \cite{G63,S63}
	\begin{align}
	    \label{eq:GS}
		\hat\rho=\int d^2\alpha \int d^2\beta\, P(\alpha,\beta) |\alpha\rangle\langle\alpha|\otimes |\beta\rangle\langle\beta |,
	\end{align}
	where $|\alpha\rangle$ and $|\beta\rangle$ denote classically coherent states of the harmonic oscillator \cite{S26}.
	Whenever $P$ cannot be interpreted as a classical probability density, the state of light $\hat\rho$ refers to a nonclassical one \cite{TG65,M86}.
	Since the $P$ distribution is in many cases highly singular \cite{S16}, thus experimentally inaccessible, regularization and direct sampling procedures have been proposed and implemented to reconstruct a function $P_\Omega$ which is always regular and non-negative for any classical states of light \cite{KV10,KVHS11}.
	This is achieved by a convolution of the Glauber-Sudarshan $P$ function with a suitable, non-Gaussian kernel $\Omega$, resulting in $P_\Omega$.
	Our previous theoretical studies suggest that the state in Eq. \eqref{eq:theo} indeed demonstrates nonclassical correlations \cite{ASV13}, i.e.,
	\begin{align}
	    \label{eq:Ncl}
		P_\Omega(\alpha,\beta)\stackrel{\text{ncl.}}{<}0
	\end{align}
	for some complex phase-space amplitudes $\alpha$ and $\beta$.

\section{Experimental implementation}
\begin{figure}[b]
	\includegraphics*[width=8cm]{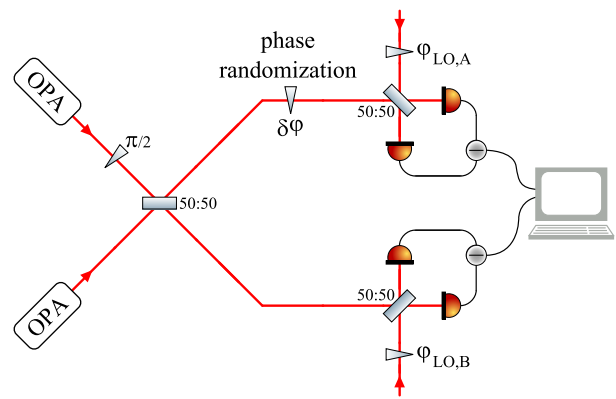}
	\caption{
		Setup outline.
		Two single-mode squeezed states are prepared by two OPAs.
		Combining these beams on a $50{:}50$ beam splitter with a $\pi/2$ phase shift yields a TMSV state.
		Introducing a phase randomization (indicated by the phase fluctuation $\delta\varphi$) in one of the output arms approximates the desired target state in Eq. \eqref{eq:theo} \cite{ASV13}.
		Both final beams are probed by balanced homodyne detectors with controllable phases $\varphi_{\mathrm{LO},A}$ and $\varphi_{\mathrm{LO},B}$.
	}\label{fig:setup}
\end{figure}

	Figure \ref{fig:setup} shows our experimental setup for the preparation and detection of the quantum state in Eq. \eqref{eq:theo}.
	The challenge here is that a full two-mode state tomography with long-term stability, phase control, and phase readout is paramount for our coherence analysis.
	See Supplemental Material \cite{SuppMat} for technical details.
	
	Two amplitude-squeezed fields at $1064\,\mathrm{nm}$ are produced by optical parametric amplifiers (OPAs).
	One OPA consists of a type-I hemilithic, standing wave, nonlinear cavity with a $7\%$ MgO:LiNbO\textsubscript{3} crystal;
	the other OPA uses a periodically poled potassium titanyl phosphate crystal instead.
	To combine distinct sources of squeezed light in one setup, the seed and pump powers are adjusted such that the squeezed output fields of the two crystals are of equal intensity and squeezing.
	For this purpose, the two pump powers of the second harmonic are chosen as $242$ and $50~\mathrm{mW}$, respectively.
	Moreover, we have to minimize power fluctuations of the pump light since the optical parametric gain of each OPA responds differently \cite{SuppMat}.

	The two squeezed fields are superposed with a visibility of $96.5\%$ and a relative phase of $\pi/2$ on a $50{:}50$ beam splitter.
	The high visibility demonstrates a successful combination of the two sources that results in a TMSV state:
	\begin{align}
	    |\mathrm{TMSV}\rangle=\frac{1}{\cosh|\xi|}\sum_{n=0}^\infty \left(
	        e^{i\arg\xi}\tanh|\xi| \right)^n |n\rangle\otimes|n\rangle.
	\end{align}

	Both output modes $A$ and $B$ of the state are probed by balanced homodyne detectors.
	We observed $(96.7\pm0.7)\%$ visibility between the fields and their corresponding local oscillators.
    For one OPA, we measured a single-mode squeezing variance of $-1.3~\mathrm{dB}$ and $+3.7~\mathrm{dB}$ antisqueezing.
	This yields an initial squeezing of $-7.3~\mathrm{dB}$ and an overall efficiency of $(63\pm2)\%$.
	The latter figure was multiplied by two to compensate for the vacuum input, because blocking the second OPA effectively introduces additional $50\%$ loss at the first (i.e., leftmost) beam splitter in Fig. \ref{fig:setup}.

    We use piezoelectric transducers to control the optical phases and realize a random phase shift $\delta\varphi$ in one of the arms.
    To achieve a uniform dephasing over the full $ 2\pi$ interval, white noise is applied with sufficiently high amplitude.
    Because of the bandwidth limitations of the transducers, a uniformly distributed phase---as required to exactly obtain the state in Eq. \eqref{eq:theo}--- can be approximated only via long measurement times, requiring to maintain a high stability of our setup.
    Further technical details are provided in the Supplemental Material \cite{SuppMat}.

\section{Results}
\begin{figure}[t]
	\includegraphics*[width=8cm]{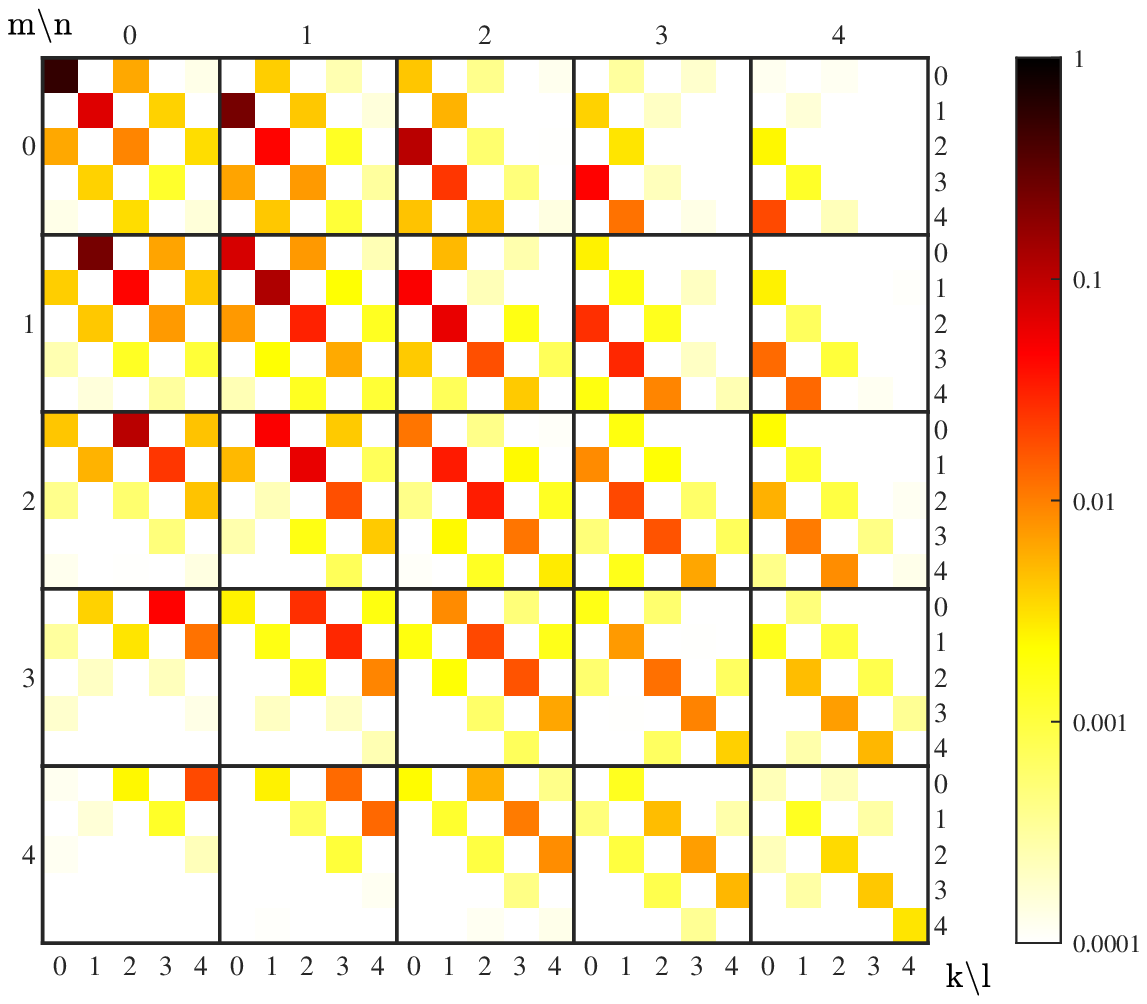}
	\includegraphics*[width=8cm]{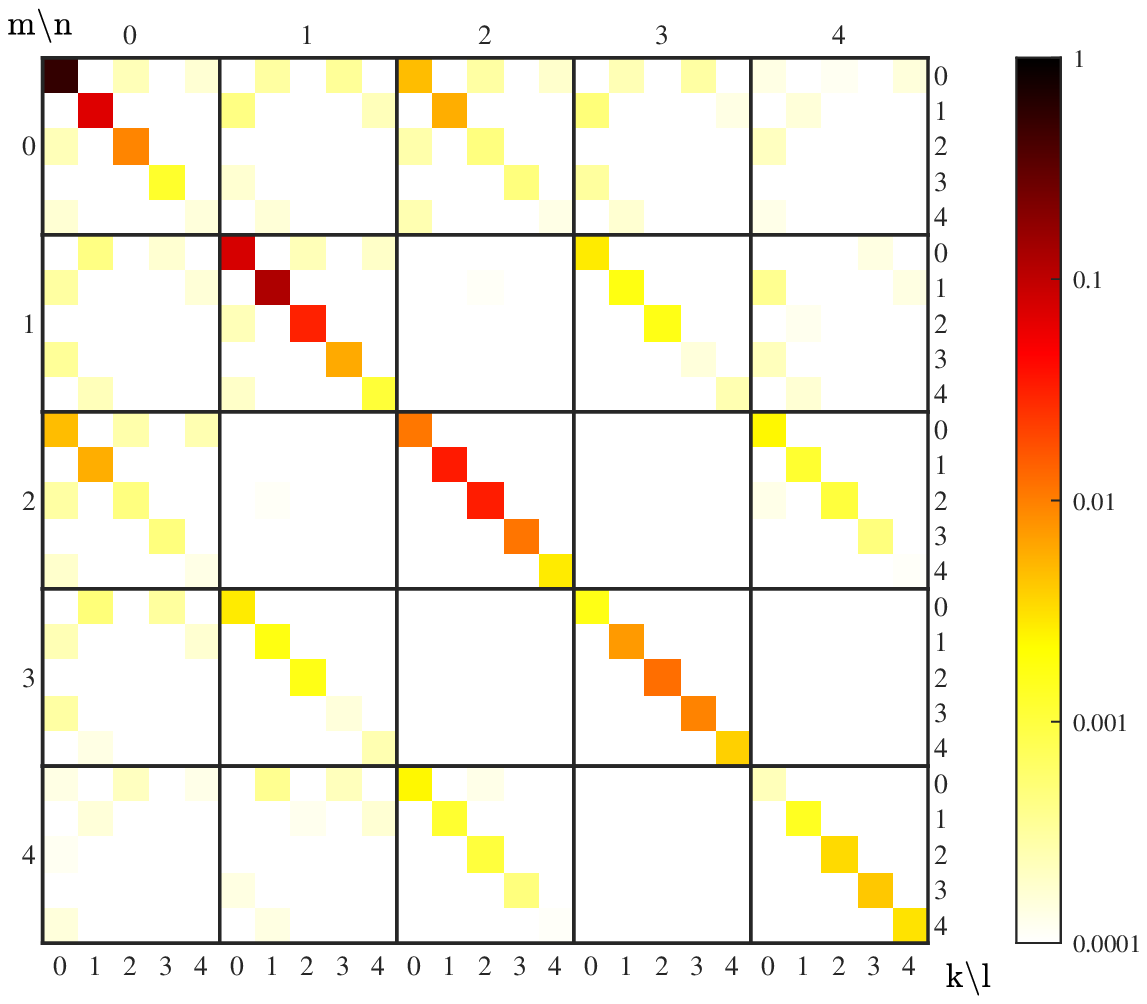}
	\caption{
		Reconstructed density matrix elements of the TMSV state in photon-number bases, before (top) and after (bottom) phase randomization.
		Each entry provides the absolute values of the density matrix elements $|\rho_{(k,m),(l,n)}|$, where $\hat\rho=\sum_{k,l,m,n} \rho_{(k,m),(l,n)} |k\rangle\langle l|\otimes |m\rangle\langle n|$.
		Please note the logarithmic scale.
		In both plots, $k$ (bottom axis) and $l$ (right axis) denote photon numbers for $A$, and $m$ (left axis) and $n$ (top axis) indicate photon numbers for $B$.
		Large off-diagonal contributions certify the presence of quantum coherence (top).
		Strongly diminished off-diagonal elements indicate the absence of quantum coherence (bottom).
	}\label{fig:dment}
\end{figure}

	Figure \ref{fig:dment} depicts the first $625$ density-matrix elements of the reconstructed state, without (top panel) and with (bottom panel) phase randomization, which requires the full two-mode state tomography.
	The initially generated state shows strong contributions of off-diagonal elements, relating to the dominance of quantum coherence [Eq. \eqref{eq:CohMeasure}], $\mathcal C(|\mathrm{TMSV}\rangle\langle \mathrm{TMSV}|)=1.789\pm0.021$.
	The implemented phase randomization then leads to a 45-fold suppression of the initial coherence, resulting in almost no subsisting coherence, $\mathcal C(\hat\rho)=0.041\pm0.005$.
	(Errors have been obtained through a Monte Carlo approach; see Supplemental Material \cite{SuppMat} for details on data processing and a discussion of the residual amount of quantum coherence, caused by experimental imperfections.)
	This loss of coherence implies that entanglement and discord do not contribute to the phase-randomized state that is almost completely characterized by its diagonal elements in the photon-number basis; see Eq. \eqref{eq:theo} and the bottom plot in Fig. \ref{fig:dment}.

\begin{figure}[t]
	\centering
	\includegraphics*[width=82mm]{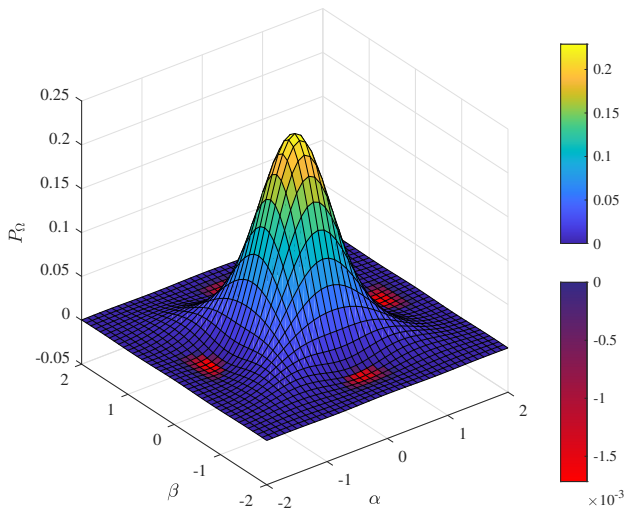}
	\caption{
		Regularized $P$ function sampled from experimental data for the phase-randomized TMSV state, including negativities.
	}\label{fig:regP}
\end{figure}

	We showed that quantum correlations in terms of quantum coherence are negligible for the produced state.
	That is, the phase-averaged state is mostly consistent with a classical statistical mixture of orthonormal two-mode, tensor-product, photon-number states, thus also ruling out entanglement and discord as a source of quantum correlations as discussed earlier.
    However, from our balanced homodyne detection data, we can further directly sample the regularized phase-space function $P_\Omega$ \cite{KVHS11}.
    Here, the reconstruction is based on phase-insensitive pattern functions \cite{SuppMat}.
    Thus, no amount of residual coherence can contribute to nonclassicality in the form of Eq. \eqref{eq:Ncl}.
	The resulting distribution is shown in Fig. \ref{fig:regP}.
    We found a maximal statistical significance of more than 150 standard deviations for the negativity of the reconstructed quasiprobability distribution, $P_\Omega(\alpha=0,\beta=1.5)=(-1.570\pm0.010)\times 10^{-3}$.
	Because of these highly significant negativities, quantum correlations in the generated state are confirmed beyond the previously considered notions.

	Again, we emphasize that quantum coherence, entanglement, and discord cannot contribute to the negativity as our approach is, by construction, insensitive to such phase-sensitive phenomena.
	Furthermore, our method is, to our knowledge, the only existing approach to uncover this form of quantumness.
	As mentioned before, the Wigner function, for instance, is completely positive for our state.
	Thus, we experimentally generated quantum correlations which are inaccessible via other means.
	Moreover, phase stability is not required for the kind of quantum effect.
	In fact, we artificially introduced phase noise---which is often omnipresent in realistic quantum channels---to produce the sought-after state.

\section{Entanglement activation}
	Because of the ever-growing importance for quantum information processing \cite{HHHH09,NC00}, the question arises if entanglement can be activated, like for coherence and discord \cite{SSDBA15,KSP16,MYGVG16}, from the nonclassically correlated state under study.
	Achieving such an activation renders this state a useful resource for quantum technologies.

	For this purpose, let us recall that single-mode nonclassicality can be converted into entanglement via simple beam splitters \cite{X02,KSBK02,VS14}.
	Similarly, we consider combining each of our two modes separately on a 50:50 beam splitter with vacuum, where annihilation operators for the nonvacuum input map as $\hat a\mapsto(\hat a+\hat a')/\sqrt 2$ and $\hat b\mapsto(\hat b+\hat b')/\sqrt 2$;
	additional modes obtained from the splitting are indicated by prime superscripts.
	Such operations are free (i.e., classical) ones with respect to the reference $|\alpha\rangle\otimes|\alpha'\rangle$ of the Glauber-Sudarshan representation [Eq. \eqref{eq:GS}] since the beam-splitter output for mode $A$ remains in this family of states, $|(\alpha+\alpha')/\sqrt{2}\rangle\otimes|(\alpha-\alpha')/\sqrt 2\rangle$, and likewise for $B$.
	Furthermore, applied to a photon-number input state $|n\rangle\otimes |0\rangle$, the map yields the output state $|\Psi_n\rangle=2^{-n/2}\sum_{j=0}^n \binom{n}{j}^{1/2} (-1)^{n-j} |j\rangle\otimes|n-j\rangle'$.
	Therefore, the state in Eq. \eqref{eq:theo} results in the final four-mode state
	\begin{align}
		\label{eq:4mode}
		\hat\rho_{AA'BB'}=\sum_{n\in\mathbb N} (1-p)p^n |\Psi_n\rangle\langle\Psi_n|\otimes|\Psi_n\rangle\langle\Psi_n|.
	\end{align}

	Clearly, this state is still nonentangled when separating the joint subsystems $AA'$ and $BB'$ from each other.
	However, multimode entanglement is much richer since entanglement in various mode decompositions can be considered;
	see, e.g., Refs. \cite{GSVCRTF15,GSVCRTF16}.
	Here, we address the question whether there is entanglement between the primed and unprimed modes, i.e., with respect to the separation of $AB$ and $A'B'$.

	To answer this question, we consider the partial transposition criterion \cite{P96,HHH96}.
	In one form \cite{HHH96}, this criterion states that a state is entangled if the expectation value of a so-called entanglement witness $\hat W=(|\Phi\rangle\langle\Phi|)^\mathrm{PT'}$ is negative, where $\mathrm{PT'}$ denotes the partial transposition of the primed modes.
	For example, we can choose $|\Phi\rangle=|0\rangle\otimes|0\rangle'\otimes|1\rangle\otimes|1\rangle'-|1\rangle\otimes|1\rangle'\otimes|0\rangle\otimes|0\rangle'$, which results in
	\begin{align}
		\mathrm{tr}(\hat W\hat\rho_{AA'BB'})=-\frac{(1-p)p}{2}<0,
	\end{align}
	for all nontrivial parameters $0<p<1$ that define our states in Eq. \eqref{eq:theo}.
	Therefore, the two-mode nonclassical correlations of the generated state can be successfully activated to produce four-mode entanglement, which is a useful mixed-state quantum resource, such as for teleportation \cite{P94,G96,HH96,HHH01}, and which can be further enhanced via entanglement distillation \cite{BBPSSW96,BBPS96,PSBZ01,KBSG01,HSDFFS08}.

\section{Conclusion}
	We experimentally realized a quantum state with quantum correlations that are inaccessible by means of two-mode coherence, thus entanglement and discord, but can be intuitively visualized by negative quasiprobabilities.
	Quantum coherence---a recently explored resource for quantum information processing---exists in terms of superpositions of orthogonal photon-number states of the initial TMSV state.
	But a phase averaging destroys this and related kinds of quantum correlations.
	Thus, contemporary quantum-information-based concepts of correlation fail to certify the quantumness of the state under the challenging, but common scenario of dephasing.
	However, the Glauber-Sudarshan-based concept of nonclassiality of light---frequently considered to be a dated notion, or not considered at all--- is still capable of uncovering the quantum nature of the generated state.

	Since the Glauber-Sudarshan distribution often displays a highly singular behavior, we employ a technique which enables us to directly sample a regularized version of such a two-mode phase-space function, allowing us to certify nonclassical negativities with a statistical significance of more than 150 standard deviations.
	Thus, the generated phase-independent state exhibits nonclassical correlations beyond entanglement and discord.
	Furthermore, we theoretically devised a method to activate multimode entanglement, utilizing only the produced state and simple beam splitter operations.

	In conclusion, we realized a type of quantum correlation that can be accessed via phase-space approaches but not through more recent notions of quantumness.
	This kind of correlation is intrinsically robust against dephasing and can be easily converted into multimode entanglement.
	Therefore, this finding offers useful quantum correlations that can be realized without experimentally costly phase stabilization and is resourceful for modern quantum information protocols.

\section*{Acknowledgements}
	E. A. acknowledges funding from the European Union's Horizon 2020 research and innovation program under the Marie Sk\l{}odowska-Curie IF InDiQE (EU Project No. 845486).
	M. S. acknowledges financial support by the Deutsche Forschungsgemeinschaft through STA-543/9-1.
	J. S. thanks Torsten Meier and Tim Bartley for discussions.
    This work was supported by the Deutsche Forschungsgemeinschaft through SFB 652, Projects No. B12 and No. B13.

\appendix

\section*{Supplemental Material}

	In this supplementary document, we provide additional details on the experiment and data processing.
	The experimental implementation and data handling for phase readout and randomization are considered in Appendices \ref{appsec:details} and \ref{appsec:phase}, respectively.
	Aspects of the reconstruction of the density matrix in the photon-number basis are provided in Appendix \ref{appsec:Fock}.
	Eventually, in Appendix \ref{appsec:P}, technical details on the reconstruction of the phase-space representation are discussed.

\section{Experimental details}\label{appsec:details}

\begin{figure*}
	\centering
	\includegraphics*[width=\textwidth]{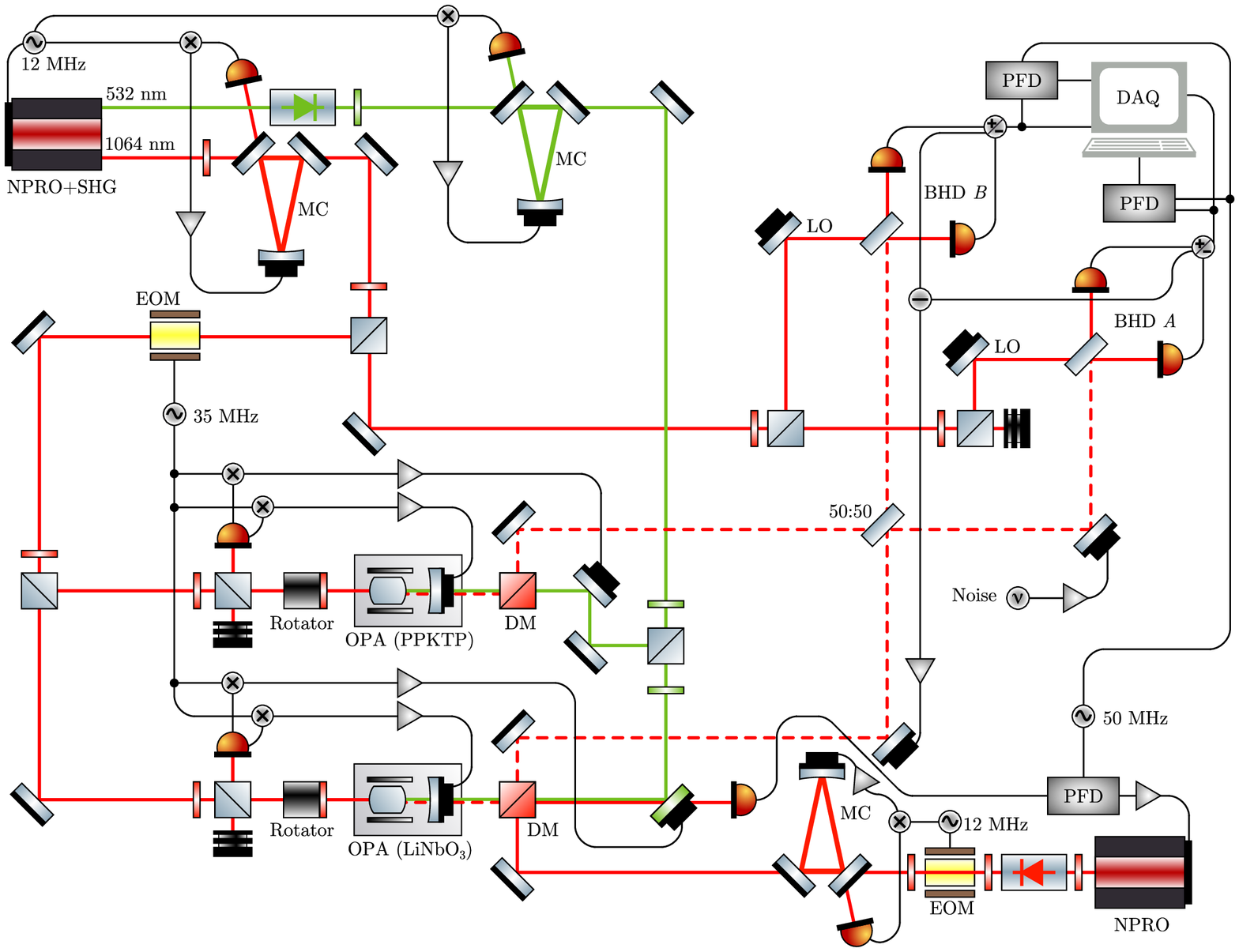}
	\caption{%
		Detailed sketch of our setup.
		BHD: balanced homodyne detector;
		DAQ: data acquisition;
		DM: dichroic mirror;
		EOM: electro-optic modulator;
		LO: local oscillator;
		MC: mode cleaner;
		NPRO: nonplanar ring oscillator;
		OPA: optical parametric amplifier;
		PFD: phase-frequency detector;
		PPKTP: periodically poled potassium titanyl phosphate;
		SHG: second-harmonic generation.
	}\label{fig:setup-detailed}
\end{figure*}

    The experimental layout is depicted in Fig. \ref{fig:setup-detailed} and is described in detail below.

    The used laser is a monolithic, nonplanar ring oscillator (NPRO), whose geometry enables a highly stable laser frequency and delivers a narrow linewidth of $1~\mathrm{kHz}$.
    A noise eater is integrated in the laser too, suppressing fluctuations in the laser intensity.
    The laser light is filtered using custom-built mode cleaners (MC), operated at a finesse of 422.
    We obtain a shot-noise-limited laser beam with a pure spatial laser profile.
    This is essential as the experimental results are extremely sensitive to noise and mode-matching imperfections.

    The heart of the setup comprises two distinct optical parametric amplifiers (OPAs).
    A challenge that we have overcome is to combine both OPAs.
    For our experiment, which ideally demands two identically squeezed states, the cavity with a MgO:LiNbO\textsubscript{3} crystal requires a much higher pump power of $242~\mathrm{mW}$ than the periodically poled potassium titanyl phosphate crystal (PPKTP) with $50~\mathrm{mW}$ to obtain the same level of parametric gain.
    It is particularly demanding that power fluctuations of the pump light do not lead to the same change of the optical parametric gain in both OPAs.
    Thus, the high stability of the laser source and control loops is paramount.

    The OPAs emit pairs of photons with frequencies $\omega \pm \Omega$, two different optical frequencies centered around the fundamental frequency.
    Frequency components of the measured signal correspond to both fields.
    Hence, we characterize the quantum state by looking at the noise in the sidebands of our signals.
    The light fields emerging from the OPAs also have a coherent amplitude because of the injected seed beam.
    This amplitude makes it easier to stabilize the cavities and measure quantities such as the optical parametric gain.
    Nevertheless, this amplitude does not significantly contribute to the quantum state of the sideband mode, but it causes modulation frequencies of the electro-optic modulators (EOMs) to occur in the signals of the balanced homodyne detectors (BHDs).
    Other interfering signals, especially low-frequency ones, are also present, which are accounted for by limiting the analysis of quantum noise to parts of the spectrum that are free of technical noise.
    In our case, we selected a range between $4$ and $8~\mathrm{Mhz}$ in postprocessing by applying eighth-order Chebyshev filters.

    The two generated single-mode squeezed beams from both OPAs are superimposed on a $50{:}50$ beam splitter.
    The desired phase relation of $\pi/2$ for two-mode squeezed vacuum (TMSV) is accompanied by equal light powers at the beam splitter outputs.
    Therefore, the difference between the total powers of BHD $A$ and BHD $B$ is chosen as the control signal.
    Electronic adders and a subtractor are used for this purpose.

    To read the optical phase at the BHDs, we use a second NPRO that is part of a phase-locked loop. 
    This loop consists of a custom-built phase-frequency detector (PFD), a controller, and a controllable oscillator.
    The latter yields the beat signal between the light fields of the two laser systems, which is caused by the superposition at the dichroic mirror (DM).
    A photodetector converts the oscillating photocurrent into a voltage signal, which is compared with the reference oscillator ($50~\mathrm{MHz}$) via the PFD.
    Deviations in frequency and phase are passed to the controller as an error signal.
    To close the control loop, a control signal is applied to a piezo actuator in the second laser, which changes the resonator length and thus detunes the laser frequency correspondingly.
    With two additional PFDs at the outputs of the BHDs, it is then possible to read out the optical phase.
    The PFDs yield the differential phase between local oscillator (LO) and the reference oscillator, which, in turn, has a fixed phase relation to the signal field due to the control loop.

    As emphasized above, a high stability is required.
    Beyond that, this stability must be maintained for a long period of time in order to be able to record the amount of data that is required for the analysis.
    For the calculation of the density matrices, the phases measured in the BHDs are each assigned to intervals of $12^\circ$.
    This gives a total of $(360/12)^2=900$ phase combinations from both BHDs.
    For each one of these settings, the TMSV state are phase-averaged by using piezo noise.
    The longer the measurement, the closer we approach the required uniform phase noise.
    The total measurement time in our case was about thirty minutes.

    In order to characterize squeezing and losses, as specified in the Letter, additional measurements were carried out in which one OPA was blocked at a time.
    In addition to the parameters given in the main text, we obtain the following relevant system parameters: 
    the signal power of the OPA's outputs is $0.030~\mathrm{mW}$;
    the powers of the LOs are $0.950~\mathrm{mW}$ each;
    and the parametric gain and suppression (anti-gain) are $2.86$ and $0.50$, respectively.

\section{Phase readout and randomization}\label{appsec:phase}

    The density matrices were calculated from quadrature histograms using pattern functions for the photon-number basis expansion \cite{ALP95,LMKRR96,R96}.
    For this reconstruction, the quadrature data needed to be allocated to specific optical phases $\varphi_{\mathrm{LO},A}$ and $\varphi_{\mathrm{LO},B}$ for modes $A$ and $B$, respectively;
    see also Fig. 1 in the main text.
    As described above, the phase range from $0$ to $2\pi$ was divided into $30$ equidistant parts, resulting in $900$ histograms or phase combinations between $A$ and $B$.
    To obtain the current optical phase within the balanced homodyne detectors, an auxiliary laser was coupled to the output field of an OPA, creating a phase-locked loop.
    Accordingly, optical phases could be resolved by feeding the AC signals of the homodyne detectors into phase frequency detectors with same beat frequency.
    See previous section for further details.

    The optical phase in arm $A$ was randomized within our experiment.
    For this purpose, a piezoelectric transducer was supplied with a low pass filtered white noise signal to match the constant part of the piezo's transfer function.
    For small amplitudes, the phase fluctuates around the preset reference phase.
	With a sufficiently high amplitude, however, the phase fluctuates over several periods and becomes uniformly distributed.
    From simulations, we found that the standard deviation of the piezo movement needed to be larger than $3.7~\mathrm{rad}$ to approach uniformly distributed phases.
    Due to the bandwidth limitations of the transducers, a uniformly distributed phase is only achievable by measuring infinitely long.
    Therefore, even the smallest density matrix elements are expected to have a value differing from zero.

    In contrast to the continuous phase recording in mode $B$, the optical phase at $A$ was fixed for time intervals of $120~\mathrm{ms}$, and the phase value determined when the noise signal was zero.
    All data recorded during this time were allocated to that reference phase.
    Long-term drifts contribute to a change in the reference phase at $A$.
    However, in addition, we applied a changing offset signal every time a new reference phase was set that allowed us to collect similar numbers of data points for each phase.

\section{Impurities in the suppression of off-diagonal elements and statistical analysis}\label{appsec:Fock}

    \begin{figure}[b]
	\centering
	\includegraphics*[width=\columnwidth]{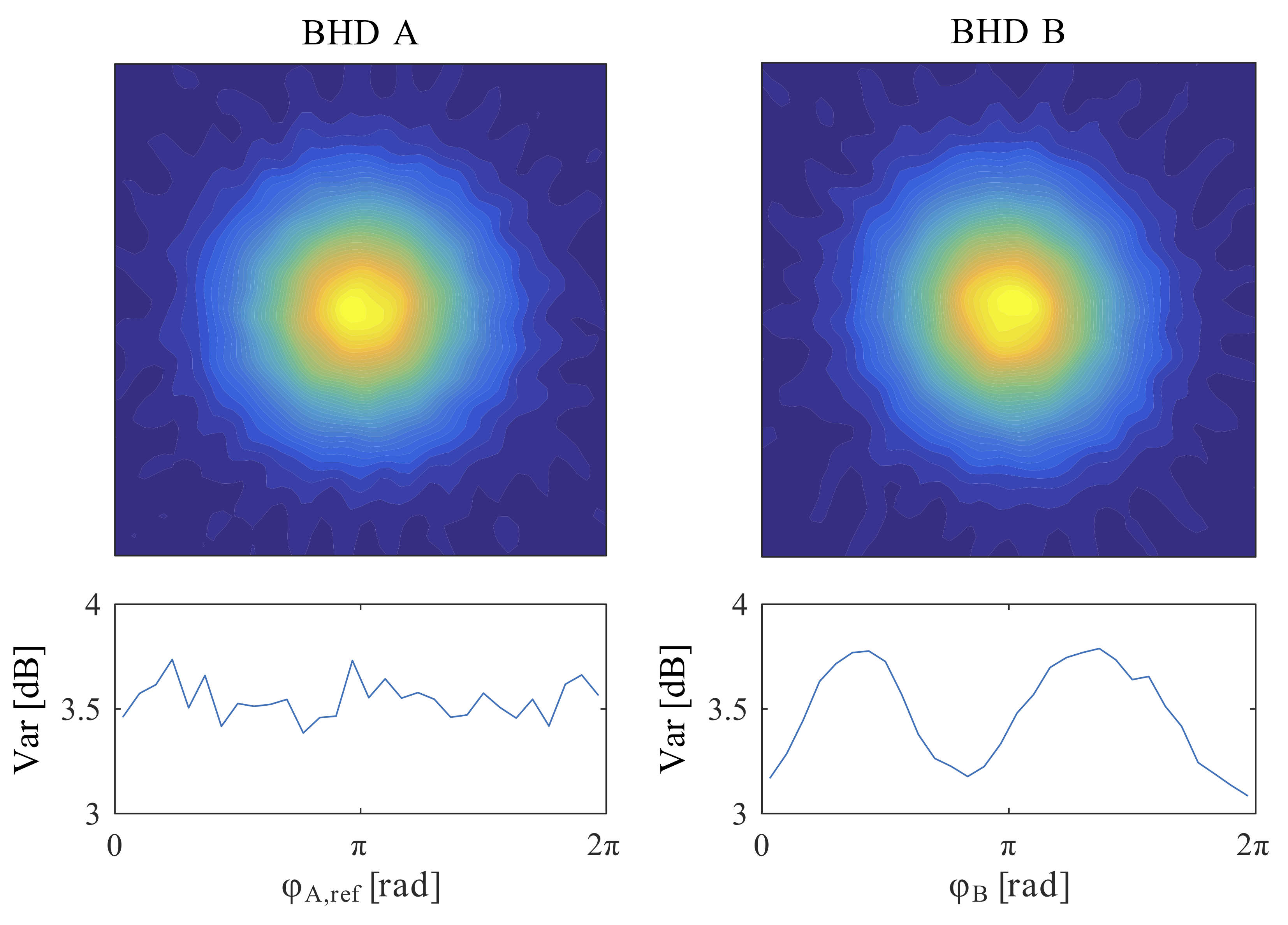}
	\caption{%
		Reconstructed marginal Wigner functions for modes $A$ and $B$, viewed as density plots (top) and quadrature variances normalized to quantum noise limit (bottom).
		Note that only a fraction of all measurement data from balanced homodyne detection (BHD) was used for this visualization.
		The left plot for mode $A$ is almost perfectly circular as one would expect for the marginal thermal state of the phase-averaged TMSV state under study.
		The right plot shows asymmetries, i.e., a slightly visible eccentricity in the Wigner function (top) and phase dependence in the variances (bottom).
	}\label{fig:wigner}
\end{figure}

    As one can see in Fig. 2 of the main text, all off-diagonal elements in the photon-number basis of subsystem $A$ approach the expected value of zero.
    In the basis of subsystem $B$, small nonzero off-diagonal elements remain.
    This may be unexpected as phase randomization in one subsystem should be sufficient for extinction of off-diagonal elements in both subsystems in theory.
    In the experimental data, these nonvanishing entries can be traced back to having not exactly the same amount of squeezing from the two distinct OPAs and potentially not having exactly $\pi/2$ phase shift between the two single-mode squeezed states at the first beam splitter (cf. Fig. 1 in the main text).
    This also results in slightly different Wigner functions and quadrature variances, as depicted in Fig. \ref{fig:wigner}, where mode $B$ displays a residual amount of phase dependence.

	Additional imperfections are caused by deviations from a perfectly uniform distribution of phases because of bandwidth limitations for implementing white noise with our transducers; see previous section.

\begin{figure}[t]
	\includegraphics*[width=\columnwidth]{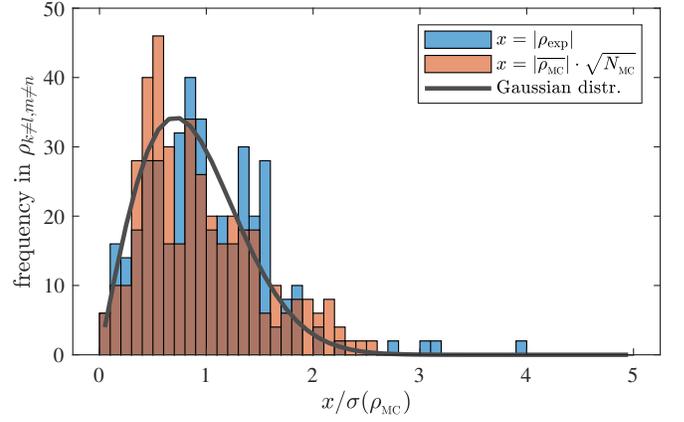} 
	\caption{%
		Histogram for off-diagonal elements of the reconstructed density matrix of the fully phase-averaged TMSV state. 
		The blue bars represent the experimental data.
		Orange bars originate from a Monte Carlo simulation.
		Both experimental and simulation histograms are normalized to standard deviations obtained from the Monte Carlo simulation.
	}\label{fig:histoffdiag}
\end{figure}

	Further deviations from zero are due to inherent random errors.
	To show that the deviations from zero are mostly of statistical nature, we performed a Monte Carlo simulation using our experimental parameters.
	This allowed us to determine the standard deviation of each element in the density matrix.
	Furthermore, the simulation also shows that, with increasing amount of the quadrature data ($N$) used to calculate the density matrix, the absolute values of the off-diagonal density matrix elements decrease accordingly, i.e., by the factor $\sqrt{N}$.
	In Fig. \ref{fig:histoffdiag}, we show in a histogram the frequency of off-diagonal density matrix elements, normalized to their standard deviation.
	The distribution shows a similar shape for our experimental data and the Monte Carlo simulation in which the complex, off-diagonal elements are normally distributed around zero (solid curve in Fig. \ref{fig:histoffdiag}).
	This leads to a coherence level $\mathcal C_\mathrm{MC}=0.039$ [with an uncertainty $\sigma(\mathcal C_\mathrm{MC})=0.005$] for the state under study, which is in agreement with the estimate $\mathcal C_\mathrm{exp}=0.041$ from our photon-number basis reconstruction.

\section{Phase-space distributions and statistical analysis}\label{appsec:P}
	
	\begin{figure*}
    \centering
    \includegraphics*[width=.475\textwidth]{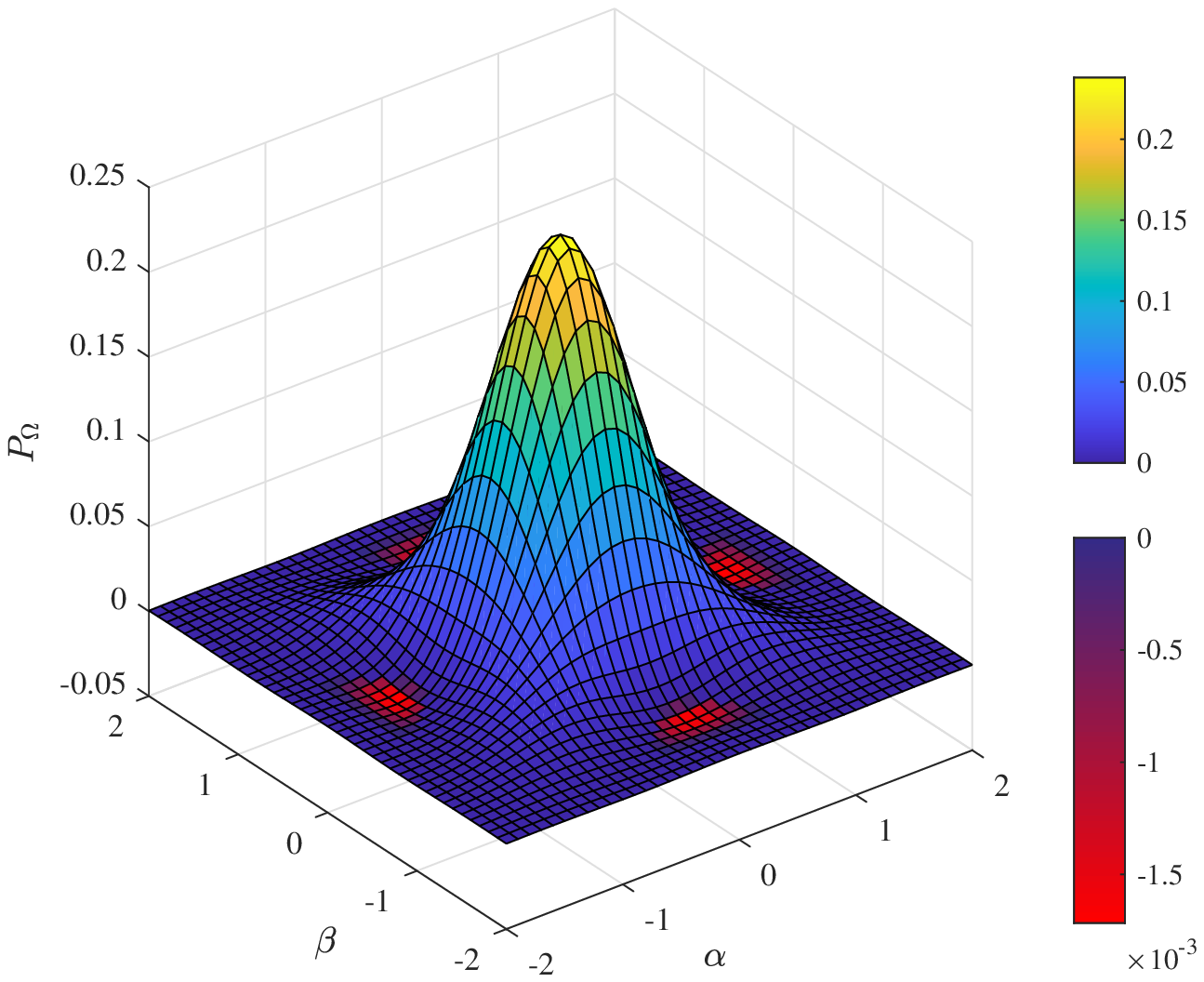}
    \hfill 
    \includegraphics*[width=.475\textwidth]{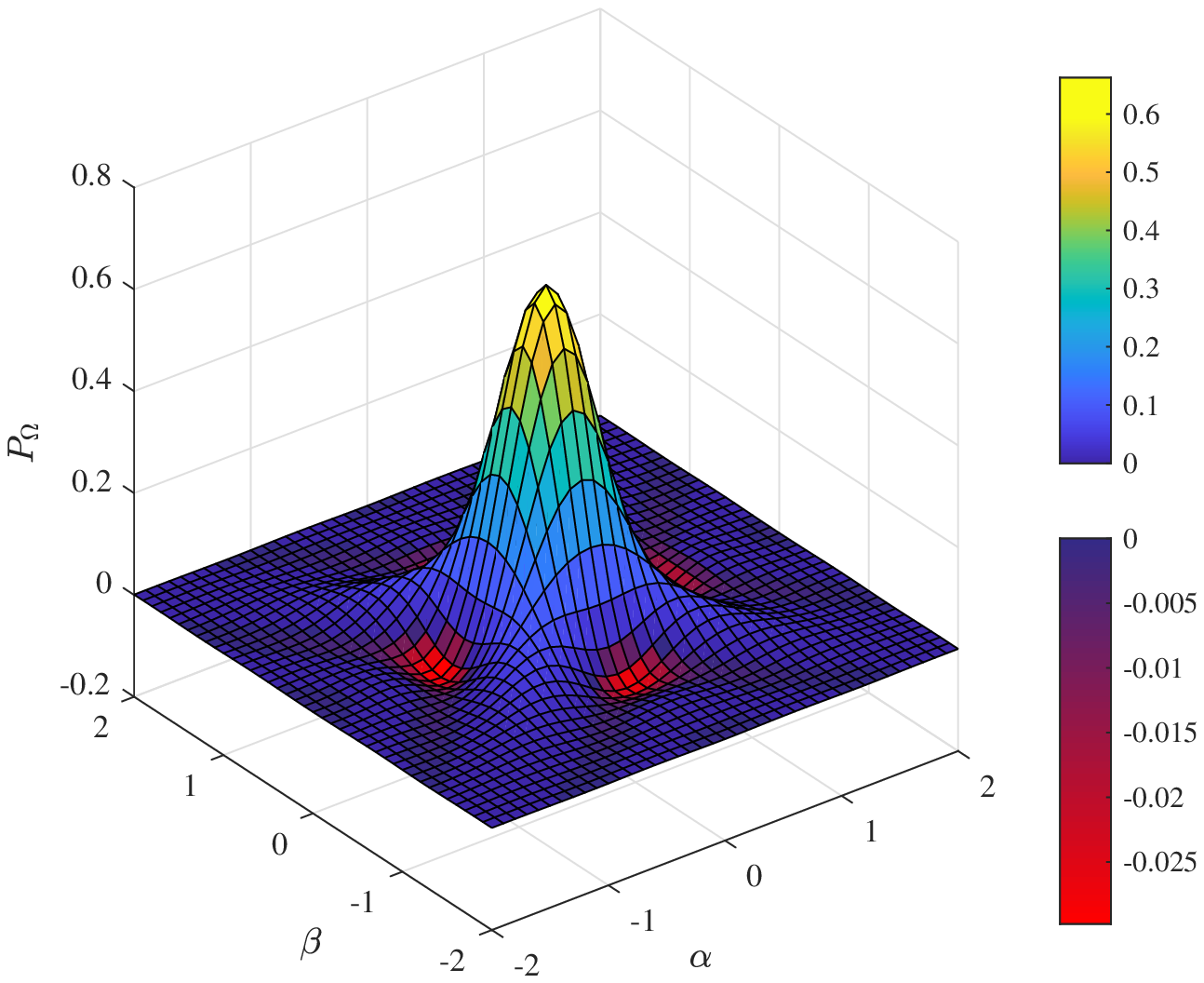}
    \caption{%
        Left:
        Regularized $P$ function for a width parameter $w=1.3$ from simulated data for a perfectly phase-randomized TMSV state with two identical squeezing sources and similar parameters as in the experiment ($-7.4~\mathrm{dB}$ squeezing, $60\%$ quantum efficiency).
        Right:
        Regularized $P$ function, sampled from experimental data for the phase-randomized TMSV state for a different width parameter, $w=1.6$.
        It shows more pronounced but less statistically significant negativities than for the optimal value $w=1.3$.
    }\label{fig:otherPs}
\end{figure*}

    For obtaining our regularized phase-space quasiprobabilities, we closely follow the approach in Ref. \cite{KVHS11} and its multimode generalization in Ref. \cite{ASV13}.
    That is, we have
    \begin{equation}
	    P_{\Omega}(\alpha,\beta){=}
	    \frac{1}{N}
	    \sum_{j=1}^{N}
	    f\left(x_A^{(j)},\varphi_{\mathrm{LO},A}^{(j)},\alpha;w\right)
	    f\left(x_B^{(j)},\varphi_{\mathrm{LO},B}^{(j)},\beta;w\right)\!,
	\end{equation}
	where $j=1,\ldots,N$ lists the measured two-mode quadrature data points $(x_A^{(j)},x_B^{(j)})$ for the phase combination $(\varphi_{\mathrm{LO},A}^{(j)},\varphi_{\mathrm{LO},B}^{(j)})$.
	Analogously to the pattern function of the photon-number-basis reconstruction, $f$ is the corresponding pattern function of the regularized $P_\Omega$, including the width parameter $w$ that is discussed below.

    In Refs. \cite{KVHS11,ASV13}, this pattern function takes the form
    \begin{equation}
        \label{eq:SamplingFormula}
    \begin{aligned}
	    & f(x,\varphi,\alpha;w)
	    \\
	    = & \frac{1}{\pi} \int\limits_{-\infty}^{\infty} dz\,|z|\, e^{z^2/2}\, \tilde\Omega(z/w)
	    \\
	    & \times
	    \exp\left[
	        izx+2i|\alpha|z
	        \sin\left(
	            \arg\alpha-\varphi-\frac{\pi}{2}
	        \right)
	    \right]
    \end{aligned}
	\end{equation}
	for a so-called filter function $\tilde\Omega(\gamma)$, being the autocorrelation function $\tilde\Omega(\gamma)=(2/\pi)^{3/2} \int d^2\gamma ' e^{-|\gamma+\gamma '|^4}\,e^{-|\gamma '|^4}$ \cite{KV10} and the Fourier transform of the positive kernel $\Omega$ mentioned in the main text.
	Here, however, we apply a phase-averaged version to ensure that negativities in the regularized $P$ function are not a result of any residual phase-dependence from an incomplete phase-averaging of the TMSV state.
	That is, we utilize the pattern function
	\begin{equation}
	\begin{aligned}
        &\bar f(x,\alpha;w) =
        \frac{1}{2\pi}\int\limits_0^{2\pi}d\varphi\, f(x,\varphi,\alpha;w)
        \\
        = & \frac{1}{\pi} \int\limits_{-\infty}^{\infty} dz\, |z|\, e^{z^2/2}\, \tilde\Omega(z/w)\, \exp[izx]\, J_0(2 z | \alpha |),
    \end{aligned}	 
	\end{equation}
    where $J_0$ is the zeroth Bessel function of the first kind, instead and sample--- according to Eq. \eqref{eq:SamplingFormula}---our regularized two-mode quasiprobabilities with this $\bar f$ rather than $f$.
    
    Thereby, and in terms of the original Glauber-Sudarshan representation, we obtain a fully phase-averaged version of the state under consideration,
    \begin{equation}
    \label{eq:PhaseAveragedState}
    \begin{aligned}
        \overline{\hat\rho}=&\int d^2\alpha\int d^2\beta
        \int\limits_0^{2\pi}\frac{d\varphi_A}{2\pi}
        \int\limits_0^{2\pi}\frac{d\varphi_B}{2\pi}
        \\
        & \times
        P(\alpha e^{i\varphi_A},\beta e^{i\varphi_B})\,
        |\alpha\rangle\langle \alpha|\otimes|\beta\rangle\langle\beta|
        \\
        =&\sum_{m,k=0}^\infty p_{m,k} |m\rangle\langle m|\otimes|k\rangle\langle k|.
    \end{aligned}
    \end{equation}
    Since phase-independent Glauber-Sudarshan $P$ distributions necessarily yield a state that is diagonal in the photon-number basis, the result in Eq. \eqref{eq:PhaseAveragedState} does not carry any information about any residual coherence whatsoever.
    In addition, the left plot in Fig. \ref{fig:otherPs} shows the resulting $P_\Omega$ for simulated data that correspond to our experiments, yet for a perfect phase randomization of the TMSV state.
    The depicted result is virtually indistinguishable from Fig. 3 in the main text.

\begin{figure}[t]
	\centering
	\includegraphics*[width=.9\columnwidth]{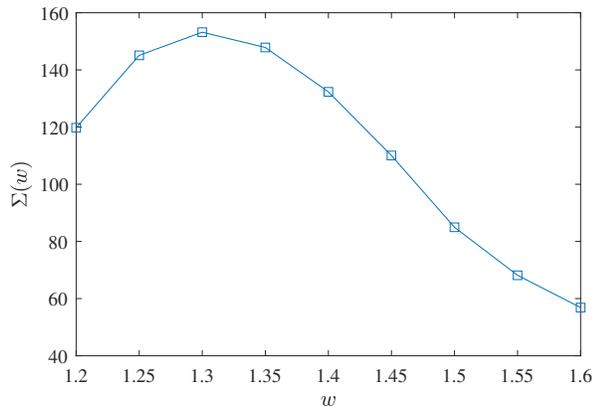}
	\caption{%
		Significance $\Sigma$ of the negativity in the regularized phase-space distribution $P_{\Omega}$ as a function of the kernel width parameter $w$.
	}\label{fig:significancew}
\end{figure}

	In order to make a statistically sound statement about the significance of the negativities in the regularized $P$ function, we divided our measured data into $N=141$ ensembles, $4\times 10^7$ quadrature data points each, and calculated $P_\Omega$ for each ensemble separately.
	The standard error of the mean can be expressed as $\sigma_N = \sigma[P_\Omega (\alpha,\beta)]/\sqrt{N}$, where $\sigma[P_\Omega (\alpha,\beta)]$ is the standard deviation for the ensembles.
	The maximal significance of the negativities $\Sigma $ for any points in phase space is
	$\Sigma=\max_{\alpha, \beta} [- P_\Omega (\alpha,\beta)/\sigma_N]$.

    As mentioned before, the regularized $P$ function is characterized by a width parameter $w$.
	For characterizing the nonclassical features of the reconstructed quasiprobability distribution $P_\Omega$, one may choose a suitable $w$ such that $\Sigma$ becomes maximal.
	For such an optimization, we found the maximum $\Sigma=153$ for $w=1.3$ (see Fig. \ref{fig:significancew}), which is used for Fig. 3 in the main text.
    Moreover, the right plot in Fig. \ref{fig:otherPs} shows $P_\Omega$ as obtained from our data for a nonoptimal width parameter.
    While the absolute value of the negativity is one order of magnitude higher here ($\approx 3\times10^{-2}$) compared with the plot in the Letter ($\approx 1.5\times10^{-3}$), the statistical significance of the certification of nonclassicality is less than half as significant than for the optimal case; see Fig. \ref{fig:significancew} (right plot) for $w=1.6$.
    Note that this trend carries on; i.e., we can increase the absolute negativity by increasing $w$ on the expense of a reduced statistical significance $\Sigma$.

\end{document}